\documentclass{article}

    \PassOptionsToPackage{numbers, compress}{natbib}
    \usepackage[final]{neurips_2023}

\usepackage[utf8]{inputenc} %
\usepackage[T1]{fontenc}    %
\usepackage{url}            %

\usepackage{breakurl}
\usepackage[backref=page]{hyperref}       %
\hypersetup{breaklinks=true}
\usepackage{booktabs}       %
\usepackage{amsfonts}       %
\usepackage{nicefrac}       %
\usepackage{microtype}      %
\usepackage{xcolor}         %
\usepackage{graphicx}
\usepackage{subcaption}
\usepackage{enumitem}
\usepackage{makecell}
\usepackage{tabularx}
\usepackage{amsmath}
\usepackage{bbm}
\usepackage{bm}

\newcommand{\reals}{\mathbb{R}}

\newcommand{\hU}{\widehat{U}}
\newcommand{\hM}{\widehat{M}}
\newcommand{\M}{\textbf{M}}

\newcommand{\cN}{\mathcal{N}}

\newtheorem{theorem}{Theorem}

\title{A Brief Tutorial on Sample Size Calculations for Fairness Audits}

\author{%
  Harvineet Singh$^{\dagger*}$,
  Fan Xia$^\dagger$, Mi-Ok Kim$^\dagger$, Romain Pirracchio$^\dagger$, Rumi Chunara$^\ddagger$, Jean Feng$^\dagger$\\
  $^\dagger$University of California, San Francisco; $^\ddagger$ New York University\\
  $^*$\texttt{fnu.harvineetsingh@ucsf.edu}
}

\begin{document}

\maketitle

\begin{abstract}
   In fairness audits, a standard objective is to detect whether a given algorithm performs substantially differently between subgroups. Properly powering the statistical analysis of such audits is crucial for obtaining informative fairness assessments, as it ensures a high probability of detecting unfairness when it exists. However, limited guidance is available on the amount of data necessary for a fairness audit, lacking directly applicable results concerning commonly used fairness metrics. Additionally, the consideration of unequal subgroup sample sizes is also missing. In this tutorial, we address these issues by providing guidance on how to determine the required subgroup sample sizes to maximize the statistical power of hypothesis tests for detecting unfairness. Our findings are applicable to audits of binary classification models and multiple fairness metrics derived as summaries of the confusion matrix. Furthermore, we discuss other aspects of audit study designs that can increase the reliability of audit results.
\end{abstract}

\section{Introduction}
Disparities in machine learning models' performance across demographic groups, subsequently referred to as unfairness, have been highlighted by audits in multiple domains including criminal justice \citep{angwin2016machine}, face image analysis \citep{buolamwini2018gendershades}, lending \citep{martinez2021mortgage}, online advertising \citep{ali2019biasedads}, hiring \citep{rhea2022personality}, and care management \citep{obermeyer2019dissecting}. Such audits have been influential to rollback products or advocate for improvements \citep{raji2022actionable}. Thus, fairness audits are an important mechanism for ensuring accountability for models. 

Frameworks exist that provide guidance on how to define, execute, and report audits of algorithmic systems \citep{raji2020closinggap,liu2022medicalaudit}. However, statistical aspects of fairness audits are largely overlooked except by a few studies \citep{yan2022auditing,chugg2023auditing,cherian2023statistical,maneriker2023refinement}. Audits produce \textit{estimates} of unfairness based on limited samples collected in the audit study. Therefore, standard considerations for making statistical inferences apply such as representativeness of the study sample to the target population, specification of the hypotheses being tested, and communicating uncertainty or statistical power of the tests. However, previous work often does not report uncertainties in the unfairness estimates (e.g. \citep{angwin2016machine,buolamwini2018gendershades}, ostensibly because the evaluation datasets used in such studies are large). Importantly, there is not enough guidance on how to design audits such that they are sufficiently powered to detect model unfairness. Previous guidances on sample sizes (e.g. \citep{riley2021validation,riley2022timetoevent,archer2021continuous}) focus on evaluation metrics such as calibration slope or area under the ROC curve which differ from the commonly-used fairness metrics such as false positive/negative rate disparity, and hence are not directly applicable.
Even though guidance exists for measures of differences between groups in the clinical trial \citep{wittes2002sample} and health disparities \citep{singh2023disparity} literature, we are not aware of works that present the formulae directly relevant to fairness metrics.

\begin{figure}[htbp!]
    \centering
    \begin{tabular}{c|cc}
       total $n_g$  &  $Y=+$   &   $Y=-$\\
       \hline
       $\widehat{Y}=+$  & $TP_g$    & $FP_g$\\
       $\widehat{Y}=-$  & $FN_g$    &   $TN_g$
    \end{tabular}
    \caption{\textbf{Confusion matrix and derived performance metrics.} The model predicts $\widehat{Y}$ for the target class $Y$ which is assumed to be either $+$ or $-$. The $n_g$ data points in a group are divided into four parts of the confusion matrix such that $n_g=TP_g+FP_g+FN_g+TN_g$ which respectively correspond to number of true positives, false positives, false negatives, and true negatives. Demographic parity, for instance, is the difference $(TP_1+FP_1)/n_1 - (TP_2+FP_2)/n_2$.}
    \label{fig:metrics}
\end{figure}

Sample size calculations are important to plan resources required to collect data since audits may not have existing data to validate the models. This is often the case in external, independent audits where the model developer might not be willing to share data (e.g. \citep{bond2021facebook}). Auditors may require conducting surveys, buying data, or querying expensive model services, thus, making data collection expensive. In some cases it is possible to leverage existing open datasets such as images of parliamentarians from across the world to audit facial image processing algorithms \citep{buolamwini2018gendershades}, data released under legal obligations by lenders \citep{martinez2021mortgage}, or direct sharing by the model developer under strict terms \citep{wilson2021auditing}. However, such datasets might not be readily available for domains such as healthcare, or we may want to evaluate the model on a new target population not measured before. Focusing on such cases, we only assume query-access to the model being audited (i.e. given an input feature vector, we can query its prediction), and optionally, the ability to collect small amount of pilot data. We assume that the goal of the audit is to conduct a hypothesis test for whether the model is fair. To this end, we describe the sample size needed such that the probability of detecting unfairness is sufficiently high. 

The main contribution of this work is to provide a tutorial on calculating sample sizes for fairness audits. We provide sample size formulae that can be directly used for auditing common fairness measures including, but not limited to, demographic (or statistical) parity \citep{calders2009demographic} and false positive/negative rates disparity \citep{chouldechova2017fair}.
By providing simple-to-use formulae, our hope is to ease the adoption and maximize the effectiveness of fairness audits.
As with any sample size calculation, the number of samples should be taken as a rough approximation. Knowing them can help to decide whether existing data is sufficient and to plan resources for the audit.

\section{Sample size calculations}
Denote the performance of the given model on a target population by $M$. This could be the accuracy, false positive rate, or different ways of measuring model performance. Suppose the population is partitioned into non-overlapping groups, such as using protected attributes like sex and race/ethnicity (or intersections thereof). Let us denote the performance for subgroup $g$ by $M_g$. Many unfairness measures are expressed as the difference in performances between two groups, i.e. $U = M_1 - M_2$ (see \citep{agarwal2018reductions} for examples). We assume that group 1 is the privileged and group 2 is the disadvantaged group, in that we believe $U$ to be non-negative. For more than two groups, one can compare performance with respect to a reference group, say, the privileged group. We restrict to binary classification models for exposition. The calculations can be extended to more than two classes.

An example of an unfairness metric is demographic parity, which is the difference between the fraction of positively predicted points between the two groups. Thus, in this case, $M_g$ is the fraction of individuals from group $g$ predicted to be positive. We will provide sample size calculations for various metrics that summarize the confusion matrix \citep{mitchell2021algorithmic} (see Figure \ref{fig:metrics} and Table \ref{tab:variance}).

Before performing any sample size calculations, we must first define the sampling design. Since we are interested in measuring group-wise performances, we need to collect enough samples from each group. One way to accomplish this is through a stratified random sampling design where individuals from each group are sampled at random with a fixed probability and independently from other individuals. There are more complex designs that might be more suited to the problem context; here we focus on stratified sampling for its applicability and simplicity. Note that for certain performance measures such as demographic parity, we do not need to collect the true labels $Y$. In general, we need the group membership, true label, and the features for computing the model predictions.

\begin{table}[htbp!]
    \caption{\textbf{Variance of performance metrics in each group.} Denote the fraction of total positives (i.e. prevalance) by $M_P=(TP+FN)/n$ and the fraction of positive predictions are denoted by $M_{PP}=(TP+FP)/n$. Denote fraction of true positives by $M_{TP}=TP/n$, true negatives $M_{TN}=TN/n$, false positives $M_{FP}=FP/n$, and false negative $M_{FN}=FN/n$. Then, true positive rate (or recall, sensitivity) is the ratio $M_{TPR}=M_{TP}/M_P$ and true negative rate (or specificity) is $M_{TNR}=M_{TN}/(1-M_P)$. We can define accuracy, false positive/negative rates and positive/negative predictive values as functions of $TP,TN,FP,FN$. All metrics are calculated on the subset of data for the group. So, $M_{TPR}$ for a group $g$ refers to the ratio $M_{TP,g}/(M_{TP,g}+M_{FN,g})$. We omit the $g$ in subscript for conciseness.}
    \label{tab:variance}
    \centering
    \begin{tabular}{ccc}
        \toprule
        Performance metric $M_g$ & Group-wise variance $\sigma^2_{M,g}$\\
        \midrule
        Demographic parity & $M_{PP}(1-M_{PP})$\\
        True positive rate (TPR) or false negative rate (FNR) & $\frac{1}{M_{P}} M_{TPR}(1-M_{TPR})$\\
        True negative rate (TNR) or false positive rate (FPR) & $\frac{1}{1-M_{P}} M_{TNR}(1-M_{TNR})$\\
        Precision/Positive predictive value (PPV) & $\frac{1}{M_{PP}}M_{TPR}(1-M_{TPR})$\\
        Negative predictive value (NPV) & $\frac{1}{1-M_{PP}}M_{TNR}(1-M_{TNR})$\\
        \bottomrule
    \end{tabular}
\end{table}

\textbf{Defining the hypothesis test.} We want to test whether the unfairness exceeds some pre-specified tolerance. That is, the null and the alternative hypotheses are
\begin{align}
\begin{split}
\text{H}_0:& U(M_1,M_2) \le U_{tol}\\
\text{H}_{alt}:& U(M_1,M_2) > U_{tol}.
\label{eq:hypo}
\end{split}
\end{align}
Here, $U_{tol}$ is the minimum tolerated unfairness which can be set to 0. We want to find the minimal sample size $n$ and allocation $(n_1,n-n_1)$ that achieves the specified power for the test at a given significance level.

The test statistic for conducting this hypothesis test is as follows.
Let $n$ denote the number of samples in the study, with $n_1,n-n_1$ samples from the two groups.
Given this, we estimate the group performances $\hM_1$ and $\hM_2$ from $n_1$ and $n-n_1$ samples respectively.
The estimate of unfairness $\hU$ is the difference between the two.
We use the test statistic as $(\hU-U_{tol})/\widehat{\sigma}_U$, where $\widehat{\sigma}_U$ is any consistent estimate of the standard deviation of unfairness (discussed later).
The null is rejected if the test statistic exceeds some threshold. 
The threshold is chosen to control the Type I error (also called significance level) at level $\alpha$ (e.g. 0.05) while maximizing power $1-\beta$ (e.g. 0.8).
Recall that Type I error is defined as falsely rejecting the null hypothesis when true (declaring a fair model as unfair) and power is defined as correctly rejecting the null hypothesis when false (declaring an unfair model as so).

The hypothesis test defined in \eqref{eq:hypo} is suitable for external audits since rejecting the null would mean that the model is unfair. 
$U_{tol}$ is specific to the problem and the performance metric. The auditor might refer to unfairness of the system prior to model deployment (such as standard of care in healthcare settings) to decide currently acceptable unfairness and set it as $U_{tol}$.
We would want high power so as to declare unfairness when present. Internal audits might prefer switching the null and alternative hypotheses to be able to conclude that the model is fair.

\textbf{Asymptotic behavior of the test statistic.} 
To find the sample size needed to power the hypothesis test, we first find the asymptotic distribution of the test statistic.
In particular, using the central limit theorem and the delta method (Theorem 3.7 in \citet{dasgupta2008asymptotic}), the estimates of the subgroup performance measures considered in Figure \ref{fig:metrics} converge to a Normal distribution with variances shown in Table \ref{tab:variance}. See Appendix \ref{sec:derivations} for details. Consequently, the test statistic converges to a Normal distribution with mean $(U-U_{tol})/\sigma_U$ and variance $1$ where $\sigma^2_U = \sigma^2_{M,1}/n_1 + \sigma^2_{M,2}/n_2$. Here, $\sigma^2_{M,g}$ are group-level variances that depend on the performance measure $M$. 

\textbf{Determining sample size and optimal allocations.} The sample size for the test needed to achieve power of $1-\beta$ while controlling Type I error at level $\alpha$ is calculated assuming unfairness is equal to some $\tau>U_{tol}$. In particular, the minimum sample size to achieve the desired Type I error and power is
\begin{equation}
    \label{eq:anyallocation}
    n = (Z_{1-\alpha/2}+Z_{1-\beta})^2(\sigma^2_{M,1}/p_1+\sigma^2_{M,2}/(1-p_1))/(\tau-U_{tol})^2
\end{equation}
where $Z$ is the inverse CDF of the standard Normal distribution and $p_1$ is the fraction of samples $(n_1/n$) we will allocate to group 1 (e.g. see \citet{wittes2002sample}). Note that the presumed level of unfairness $\tau$ should be selected based on prior knowledge about the model. The smaller the difference between $\tau$ and $U_{tol}$, the more samples we will need.

We can find the optimal allocation $p_1$ that minimizes the required sample size $n$ by solving the first-order optimality condition for \eqref{eq:anyallocation} in terms of $p_1$. The optimal allocation is $p_1=\sigma_{M,1}/(\sigma_{M,1}+\sigma_{M,2})$, which is also known as Neyman allocation \citep{neyman1938allocation} in survey sampling literature. That is, we should allocate samples in proportion to the variability of the metric in the groups. We note that the allocation is not simply to have equal number of samples per group. How much to over- (or even under-) sample the minority group depends on the variances.
Sample size under the optimal allocation is
\begin{equation}
    \label{eq:samplesize}
    n=((Z_{1-\alpha/2}+Z_{1-\beta})(\sigma_{M,1}+\sigma_{M,2})/(\tau-U_{tol}))^2.
\end{equation}

To use these sample size formulae, we need estimates of group performances to estimate the group standard deviations $\sigma_{M,g}$ using the Table \ref{tab:variance}. These can be approximated from data used to train the model or from previously-conducted audits for the model when available. We can alternatively run a small pilot study to estimate the metrics. Since we will adapt the data collection in the main study by estimates from the pilot study, this may bias our unfairness estimates. This can be avoided by pooling estimates from pilot and main studies as described in \citet{singh2023disparity}. 

We illustrate the sample size calculations on an income prediction model trained on American Community Survey data \citep{ding2021retiring} in Appendix \ref{sec:sizes}. We observe that sample sizes widely vary depending on the metric. For example, auditing for disparity in TPR requires around $1.6$ times as many samples as for demographic parity, mainly due to the increased variance of the unfairness estimate for the TPR metric.

\subsection{Other considerations}
Even before doing sample size calculations, it is important to define the target population of the audit. This is typically the population which will be subjected to the model predictions such as loan applicants from a lender's area of operation. Identifying the target population is important to ensure that the study collects a representative sample. There might be a mismatch in the target population and the sample collected due to infeasibility of randomly sampling the target population. In this case, our inferences need to stated in terms of the sampled population.

While collecting data for the audit, there are many statistical biases to keep in mind. The target labels collected during the study may differ from the way target label was defined in the model and/or at deployment. Thus, the unfairness identified in the audit might not correspond to the actual unfairness of the model. Finally, there might be non-compliance or selective responses for subjects in the audit which leads to missing data or selection bias in the unfairness estimates. These and other considerations standard in the clinical study literature (e.g. \citep{wittes2002sample}) should be kept in mind while analysing and interpreting the audit results.

\section{Conclusion} 
Independent audits of machine learning models are challenging in part because the training or validation data are typically not available to the auditors. However, reliable estimates of unfairness are important to create public awareness of model failures and demand updates to the model. We present a method to aid auditors in deciding the number of samples needed to test for unfairness. The method applies to common fairness metrics expressed as differences (or other differentiable functions) of group-wise performance metrics, for example, demographic parity, and false positive/negative rate disparities.

As further work, the method can be extended to include performance metrics such as area under the ROC curve and calibration slope. An important next step is to address practical considerations such as missing data or measurement error in group membership or labels \citep{bao2021its}. We may want to audit for a large number of groups, especially when considering intersectional identities. How do we reduce the sample sizes required in this case while sharing information across the groups? Also, if there are other types of audits being conducted (e.g. for model calibration or stability \citep{rhea2022personality}), how does one collect a single dataset so that all the different audits are properly powered? The statistical aspects of the audits that we highlight are complementary to aspects such as independence and integrity of the auditing organization, and efficient communication of the audit results to the policy makers and public. Together these may ensure that audits become a widely-adopted mechanism for keeping model developers accountable for their societal impact.

\begin{ack}
    This work was funded through a Patient-Centered Outcomes Research Institute® (PCORI®) Award (ME-2022C1-25619).
The views presented in this work are solely the responsibility of the author(s) and do not necessarily represent the views of the PCORI®, its Board of Governors or Methodology Committee.
\end{ack}

\bibliographystyle{unsrtnat}
\bibliography{disparity}

\appendix

\section{Example}
\label{sec:sizes}
We apply our method to evaluate fairness of an income prediction model trained on survey data. American Community Survey (ACS) data contains responses on education, housing, health, demography, and many other variables from a representative sample of the US households \citep{acs2023web}.

\textbf{Experiment setup.} We consider the task of predicting income level of an individual from attributes related to their education, work, and demography. Prediction target is binary (high vs low income binarized at the income threshold of $50,000$ USD). We randomly split the dataset into train (70\%) and test (30\%). We train a gradient boosting classification model on the train data, and compute the unfairness and sample sizes on the test data. We evaluate disparity in model performance between white and Black or African American population groups. In our processed data, Black or African American population group constitutes $7\%$ of the total data points (2562 out of 38869 points). True positive rate for white population is 0.79 which is $11$ percentage points higher than for the Black or African American population at 0.68. Table \ref{tab:sizes} gives an example of the sample sizes obtained from \eqref{eq:samplesize} for unfairness defined by different performance metrics.

\begin{table}[htbp!]
    \caption{\textbf{Example of sample size calculations for an income prediction model on ACS data.} We report the unfairness (its absolute value) and sample sizes needed to detect that value for different metrics. We evaluate the $n$ at optimal allocation for different unfairness metrics and for a hypothesis test at $\alpha=0.05,\beta=0.2,U_{tol}=0$. We set $\tau$ to be the actual unfairness estimate for the metric computed on the test data. Sample sizes vary a lot depending on the metric of interest. The sample size for NPV is 846,671 mainly because we required an extremely precise value of $\tau=0.003$. Note that the numbers are only meant to illustrate the sample size formulae and are not for direct use.}
    \label{tab:sizes}
    \centering
    \begin{tabular}{ccc}
        \toprule
        Performance metric $M$ & Presumed unfairness $\tau$ & Sample size $n$ \\
        \midrule
        DP & 0.09 & 855 \\
        TPR & 0.11 & 1,379 \\
        TNR & 0.03 & 6,540 \\
        PPV & 0.06 & 4,364 \\
        NPV & 0.003 & 846,671 \\
        \bottomrule
    \end{tabular}
\end{table}

Take for instance demographic parity (DP). Rate of positive predictions $M_{PP,1},M_{PP,2}$ in Black or African American and white population groups on the test set are $0.3478,0.4404$ respectively. Thus, the unfairness value is $0.093$ which we take as $\tau$. Group variances $\sigma^2_{M,1},\sigma^2_{M,2}$ from Table \ref{tab:variance} are $0.227,0.246$. Plugging these in \eqref{eq:samplesize} we get the minimum required sample size to detect $\tau$ is
$$
n \approx (Z_{1-0.05/2}+Z_{1-0.2})^2 \times (0.227^{1/2} + 0.246^{1/2})^2 / 0.093^2 \approx 855.
$$

\section{Derivations}
\label{sec:derivations}

\textbf{Asymptotic distribution of performance metrics.} We first note that for many performance metrics, $\widehat{M}_g$ are sample averages or ratios of sample averages. For example, accuracy is an average of correct predictions, $(TP+TN)/n$. This implies that $(\hM_1,\hM_2)$ follows a multivariate Normal distribution asymptotically by the central limit theorem. It has mean $M_1,M_2$ and the variance depends on the particular performance metric $M$. This follows from \citet[][Theorem 1.3.2]{fuller2009sampling} for data collected by \textit{simple random sampling} without replacement in each group independently. Throughout we ignore the finite sample correction which multiplies $(1-n/N)$ to the variance where $N$ is the population size. We assume that the sample size is negligible compared to the population size such that $n/N\to 0$. This is the case while surveying large populations.

For true positive rate (TPR), we need the joint distribution of $TP$ and $P$ which is the following,
$$
\sqrt{n}
\begin{pmatrix}
    \hM_{TP} - M_{TP}\\
    \hM_P - M_P
\end{pmatrix}
\sim \cN \begin{pmatrix}
    \begin{pmatrix}
        0\\
        0
    \end{pmatrix},
    \begin{bmatrix}
        M_{TP}(1-M_{TP}) & M_{TP}(1-M_P)\\
        M_{TP}(1-M_P) & M_P(1-M_P)
    \end{bmatrix}
\end{pmatrix}
$$

Similarly, for true negative rate (TPR), we need the joint distribution of $TN$ and $N$ which is the following,
$$
\sqrt{n}
\begin{pmatrix}
    \hM_{TN} - M_{TN}\\
    \hM_N - M_N
\end{pmatrix}
\sim \cN \begin{pmatrix}
    \begin{pmatrix}
        0\\
        0
    \end{pmatrix},
    \begin{bmatrix}
        M_{TN}(1-M_{TN}) & M_{TN}(1-M_P)\\
        M_{TN}(1-M_P) & M_P(1-M_P)
    \end{bmatrix}
\end{pmatrix}
$$

\textbf{Asymptotic distribution of ratios of performance metrics.} Now, we require the asymptotic distribution of ratios of $TP$ and $P$, or $TN$ and $N$. Distribution of differentiable functions of Normally distributed random variables can be computed by the delta method. Denote some performance metrics of interest arranged in the vector $\M$, for instance, $(M_{TP},M_{TN},M_P)$. 

\begin{theorem}[Delta method e.g. Theorem 3.7 in \citet{dasgupta2008asymptotic}]
\label{thm:delta}
Given a sequence of k-dimensional random vectors $\{\widehat{\M}_n\}$ such that $\sqrt{n}(\widehat{\M}_n - \theta)\xrightarrow{distr.} \cN_k(0, \Sigma(\theta))$. Consider a function $d: \reals^k \to \reals$ where $d$ is once-differentiable at $\theta$ and $\nabla d(\theta)$ is the gradient vector at $\theta$. Then, we have 
\begin{equation}
\sqrt{n}\left(d(\widehat{\M}_n)-d(\theta)\right) \xrightarrow{distr.} \cN\left(0, \nabla d(\theta)^\top \Sigma(\theta)\nabla d(\theta)\right)
\end{equation}
provided $\nabla d(\theta)^\top \Sigma(\theta)\nabla d(\theta)$ is positive definite.
\end{theorem}

By Theorem \ref{thm:delta}, the empirical estimate of the performance metric $d(\hM)$ follows a Normal distribution asymptotically with mean $d(\M)$ and variance $\nabla d(\M)^\top \Sigma \nabla d(\M)$.

Variance for estimated TPR which is $\hM_{TP}/\hM_{P}$ is $\frac{1}{n M_P}M_{TPR}(1-M_{TPR})$. Similarly variance for estimated TNR is $\frac{1}{n (1-T_P)}M_{TNR}(1-M_{TNR})$.

\textbf{Asymptotic distribution of unfairness metrics.} We need to compute the variance of the unfairness estimate $\hU$. By definition, unfairness is a difference, $\hM_1-\hM_2$, of statistically independent quantities since we compute the metrics separately on independently-sampled groups. Thus the asymptotic variance of unfairness estimate is the sum of variances, $\sigma^2_U = \sigma^2_{M,1}/n_1+\sigma^2_{M,2}/n_2$. We can combine the above to get variance for disparity in TPR as
$$
\sigma^2_U = \frac{1}{M_{P,1}}M_{TPR,1}(1-M_{TPR,1}) + \frac{1}{M_{P,2}}M_{TPR,2}(1-M_{TPR,2}).
$$

For the hypothesis test, we use the test statistic as $(\hU-U_{tol})/\widehat{\sigma}_U$. For a consistent estimate $\widehat{\sigma}_U$ converges to $\sigma_U$. Then, the test statistic converges to a Normal distribution with mean $(U-U_{tol})/\sigma_U$ and variance 1 by applying the Slutsky's theorem. Now we do sample size calculations for the test. 

\textbf{Sample size calculation.} Denote standard deviation under the null and alternative hypotheses by $\sigma_{U,null},\sigma_{U,alt}$ respectively. We want significance level $\alpha$ which means the following holds for the test statistic.
\begin{align*}
    \sup_{u \le U_{tol}} P\left(\frac{\hU-U_{tol}}{\sigma_{U,null}} > Z_{1-\alpha/2} \Big{|} U=u\right) &\leq \alpha\\
    \implies P\left(\frac{\hU-U_{tol}}{\sigma_{U,null}} > Z_{1-\alpha/2} \Big{|} U=U_{tol}\right) &\leq \alpha
\end{align*}
The following should hold as well for power $1-\beta$.
\begin{align*}
    \inf_{u \ge \tau} P\left(\frac{\hU-u}{\sigma_{U,alt}} > Z_{1-\alpha/2} \Big{|} U=u\right) &\ge 1 - \beta\\
    \implies P\left(\frac{\hU-U_{tol}}{\sigma_{U,alt}} > Z_{1-\alpha/2} \Big{|} U=\tau\right) &\ge 1 - \beta\\
    \implies P\left(\frac{\hU-\tau}{\sigma_{U,alt}} > Z_{1-\alpha/2} - \frac{(\tau-U_{tol})}{\sigma_{U,alt}} \Big{|} U=\tau\right) &\ge 1 - \beta
\end{align*}
Setting $Z_{1-\alpha/2} - \frac{(\tau-U_{tol})}{\sigma_{U,alt}}\leq -Z_{1-\beta}$, we have by rearranging terms that
\begin{align*}
    \sigma^2_{U,alt}&\leq \frac{(\tau-U_{tol})^2}{(Z_{1-\alpha/2}+Z_{1-\beta})^2}\\
    \iff n&\geq \frac{(Z_{1-\alpha/2}+Z_{1-\beta})^2(\sigma^2_{M,1}/p_1+\sigma^2_{M,2}/(1-p_1))}{(\tau-U_{tol})^2}
\end{align*}
We can replace $p_1$ by the optimal allocation $\sigma_{M,1}/(\sigma_{M,1}+\sigma_{M,2})$ to get the minimum sample size.

\section{Code availability}
Code to query ACS data is taken from the \texttt{folktables} packages \citep{ding2021retiring}. Scripts to compute the sample sizes will be made available at \url{https://github.com/harvineet/sample-size-fairness-audits}.

\end{document}